\documentclass[12pt,noshowpacs,nofootinbib,notitlepage,amsmath]{revtex4-2}
\usepackage{setspace}
\allowdisplaybreaks
\usepackage{graphicx,color}
\usepackage[charter]{mathdesign}
\usepackage{enumerate}
\usepackage{array}
\DeclareSymbolFontAlphabet{\mathcal}{symbols}
\DeclareSymbolFont{symbols}{OMS}{xmdcmsy}{m}{n}
\DeclareSymbolFont{largesymbols}{OMX}{cmex}{m}{n}
\SetSymbolFont{symbols}{bold}{OMS}{xmdcmsy}{b}{n}

\begin{document}

\title{\color{blue}\Large UV-complete 4-derivative scalar field theory\footnote{In Nuclear Physics B volume dedicated to the memory of Steven Weinberg.}}
\author{Bob Holdom}
\email{bob.holdom@utoronto.ca}
\affiliation{Department of Physics, University of Toronto, Toronto, Ontario, Canada  M5S 1A7\vspace{20pt}\\{\large Dedicated to the memory of Steven Weinberg}\vspace{20pt}}
\begin{abstract}
A scalar field theory with 4-derivative kinetic terms and 4-derivative cubic and quartic couplings is presented as a proxy for quantum quadratic gravity (QQG). The scalar theory is renormalizable and asymptotically free and the remaining key issue is unitarity, or more precisely positivity, just as it is in QQG. We have extended calculations for the optical theorem and for a differential cross section, both in the high energy limit, to show how positivity constrains the theory. The results also show how it is that differential cross sections can have good high energy behavior. Finally we use the scalar theory to extend the Stuckelberg theory of a massive U(1) gauge boson to a renormalizable theory of a self-interacting gauge boson.
\end{abstract}
\maketitle

\section{1977-1981}
Steven Weinberg's research during these four years was both influential and wide-ranging, and in the middle of this activity he was awarded the Nobel Prize in 1979. These four years can be described as the culmination of the classic phase of his research. By `classic' I mean the study of quantum field theories tied as closely as possible to observation and to already established theories. Weinberg presented a classic approach to ``The problem of mass'' \cite{Weinberg:1977hb} in Sept~1977, and he continued in the following year with ``Implications of dynamical symmetry breaking: An addendum'' \cite{addendum}. The next phase of his research began in Sept 1981 when he completed his first work on ``Supersymmetry at ordinary energies'' \cite{Weinberg:1981wj}. I happened to begin the Masters-PhD program at Harvard in Sept 1977 and then a few years later followed Weinberg, as my PhD supervisor, to Austin Texas, where I finished the program in Aug 1981.

At the end of 1977 and late in 1978 Weinberg completed the two highly influential works, ``A new light boson'' \cite{Weinberg:1977ma}, and ``Phenomenological Lagrangians'' \cite{Weinberg:1978kz}. At some point he suggested to me that a systematic approach to low energy QCD might be a topic to work on. I instead decided to ponder his work in the context of new physics. In this regard he did impress on me how QCD serves as a template for what is desired in a UV complete physical theory. At this stage of his thinking he was quite willing to contemplate strong interactions in extensions of the standard model, and to employ symmetries, both broken and unbroken, to extract content from such theories. This ran somewhat counter to the growing trend at the time, which was to develop perturbative extensions of the standard model. A grand unified theory is the prime example, and in 1979 he approached this subject in his usual elegant fashion, with ``Baryon- and lepton-nonconserving processes'' \cite{Weinberg:1979sa}.

In 1979 Weinberg also completed work on "Ultraviolet Divergences in Quantum Theories of Gravitation" \cite{Weinberg:1980gg}. As background, he notes the similarity between the low energy effective theory of gravity, containing an infinite number of nonrenormalizable terms, and the low energy effective theory of QCD. He also mentions another theory that is QCD-like, the renormalizable theory of gravity that includes terms quadratic in curvature. This generates a pole with an abnormal sign, which he mentions is not consistent with unitarity. But the idea of UV completeness via renormalizability and asymptotic freedom is clearly so compelling to Weinberg that he develops a suitable generalization for gravity. This is asymptotic safety, and this remains an active area of research to this day.

\section{Introduction}
We shall extend our study \cite{Holdom:2023usn} of a 4-derivative scalar field theory, having 4-derivatives both in the interaction terms and the kinetic terms, and having both cubic and quartic couplings. The theory is defined as follows in four spacetime dimensions,
\begin{align}
{\cal L}=\frac{1}{2}\partial_\mu\phi(\Box+m^2)\partial^\mu\phi+\lambda_3(\partial_\mu\phi\partial^\mu\phi)\,\Box\phi+\lambda_4(\partial_\mu\phi\partial^\mu\phi)^2.\label{e1}
\end{align}
The real scalar field $\phi(x)$ is dimensionless, as are the couplings $\lambda_3$ and $\lambda_4$. The presence of $m^2$ breaks the classical scale invariance. The theory has a shift symmetry $\phi\to\phi+c$ and this forbids other terms that could also have dimensionless couplings. Related to this symmetry is the fact that this theory is renormalizable.

This theory serves as a proxy for another renormalizable 4-derivative theory, quantum quadratic gravity (QQG) \cite{stelle}. The terms quadratic in curvature introduce 4-derivatives in both the kinetic and interaction terms. The shift symmetry of the scalar theory is playing the role of coordinate invariance of the gravity theory. The $ m^2\partial_\mu\phi\partial^\mu\phi$ term is playing the role of the Einstein term. At low energies this term dominates and the resulting low energy theory has a normal massless field with non-renormalizable interactions. For gravity this is the known quantum effective field theory description of the massless graviton.

The scalar and gravity theories are both UV complete and so can be considered at energies much higher than $m$. In the case of gravity this corresponds to ultra-Planckian energies. We shall also refer to this as the $m\to0$ limit. The unusual aspect of both theories has to do with the four derivatives, and the implications for the two theories are very similar. Since loop calculations in particular are more straightforward in the scalar theory, we shall focus on this theory here.

Because of two and four derivatives in the quadratic term, the propagator has two poles, with the massive one having an abnormal sign. The requirement for energies to be positive determines the method of quantization, and this produces a negative norm state. This result is said not to be consistent with unitarity. But this statement is not sufficiently precise. S-matrix unitarity can still be defined in the presence of negative norm states. It is defined by $S\mathbb{1}S^\dagger=\mathbb{1}$ where a generalized identity operator reflects the negative norms via the completeness relation $\mathbb{1}=\sum_X\frac{|X\rangle \langle X|}{\langle X|X \rangle}$. This S-matrix unitarity still implies conservation of probability.

S-matrix unitarity can also be established via the optical theorem, which can be directly verified in perturbation theory by simply keeping track of minus signs that are part of the theory. The LHS of the optical theorem is a forward scattering amplitude, and its calculation is affected by any abnormal-sign propagators. The RHS is a scattering process into on-shell final states, and this is affected by any negative norms among these states. Norms of states enter a cross section in the same way they enter the Born rule in quantum mechanics. It should therefore not be surprising that the LHS and RHS of the optical theorem are both affected in such a way that it remains satisfied.

Apparently what is actually meant by ``not consistent with unitarity'' is that the theory may have problems with positivity. There are certainly some abnormal minus signs appearing in calculations, but the question is whether physical quantities that should be positive, such as probabilities and cross sections, can end up being positive. This needs to be investigated. Our focus will be on the positivity constraint in the high energy limit. We shall also return to the issue of positivity in the full theory, where the question of what is or is not an asymptotic state comes into play.

\begin{figure}
\centering
\includegraphics[width=.85\linewidth]{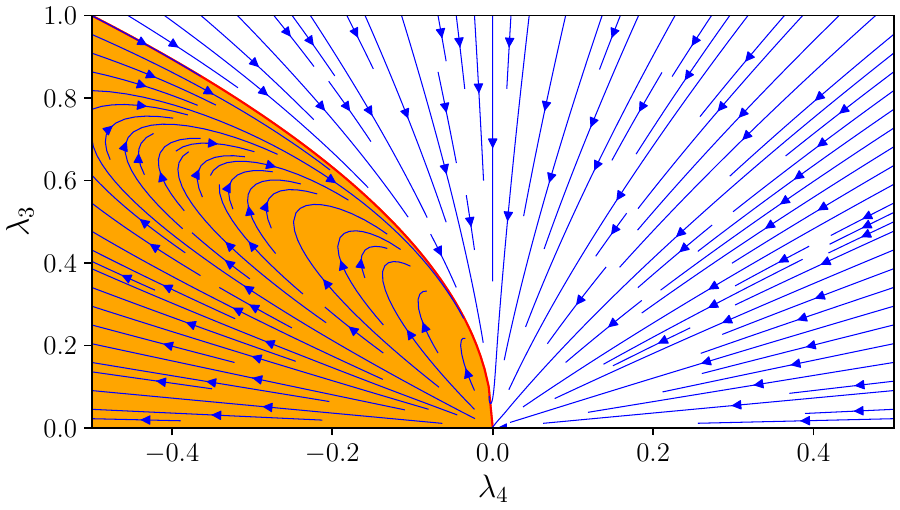}
\caption{Running coupling flow towards the UV. The flow for negative $\lambda_3$ is the mirror image of what is shown. The shading distinguishes two qualitatively different types of flows.}
\label{f5}
\end{figure}

We first look at the $\beta$-functions of the theory. They can be calculated in a standard way where the renormalization of the $\partial_\mu\phi\Box\partial^\mu\phi$ term is treated as a wave function renormalization. The result from \cite{Holdom:2023usn} is
\begin{align}
\frac{d\lambda_3}{d\ln\mu}&=-\frac{5}{4\pi^2}(\lambda_4\lambda_3+\frac{3}{4}\lambda_3^3),\\\frac{d\lambda_4}{d\ln\mu}&=-\frac{5}{4\pi^2}(\lambda_4^2+\lambda_4\lambda_3^2).
\end{align}
The resulting renormalization group flow is depicted in Fig.~\ref{f5}. The unshaded region exhibits asymptotic freedom in the UV, while the shaded region exhibits asymptotic freedom both in the UV and the IR. In this region the couplings reach maximum values at intermediate scales. The red line corresponds to $\lambda_4=-\frac{1}{2}\lambda_3^2$, and it marks the boundary between these two sets of flows that are qualitatively different.
How far the theory can proceed along a trajectory in the IR direction depends on the value of $m$. The flow towards the IR stops as the energy scale drops below $m$, since at lower energies we are left with the low energy theory mentioned above. The crossover to this low energy theory may occur at weak couplings, in which case the theory remains perturbative at all scales. But for sufficiently small $m$, and for trajectories that flow away from the origin in the IR, the couplings can continue to grow strong. A scale of strong interactions can imply the generation of a new mass scale that is above $m$, through dimensional transmutation. It is this mechanism that could be the origin of the Planck mass in QQG, since that theory also has an asymptotically free coupling.

We shall extend our work on the scalar theory \cite{Holdom:2023usn} in two ways. The first is to develop a simplified method for calculating in the high energy limit. With this method the field $\phi$ is not explicitly decomposed into two degrees of freedom, and we shall find what appears to be only one degree of freedom at high energies. The second is to find the dependence of the optical theorem on both couplings $\lambda_3$ and $\lambda_4$. We shall do the same for the differential cross section for the scattering $\phi\phi\to\phi\phi$. This will give us expressions that we can test for positivity. The requirement of positivity ends up implying a constraint on the allowed region in the space of couplings. In fact we shall find that the unshaded region in Fig.~\ref{f5} is allowed while the shaded region is not.

There is another issue with these four derivative theories, and that is due to the way 4-derivative interaction terms produce diverging tree-level amplitudes at large momenta. But this behavior for the amplitudes need not carry over to differential cross sections. Results in \cite{Holdom:2023usn} (and in \cite{Holdom:2021hlo,Holdom:2021oii} for the case of QQG) have shown that there are impressive cancellations between exclusive differential cross sections. Here we shall see that the apparent reduction to one degree of freedom helps to clarify this phenomenon. With only one degree of freedom there is just a single $2\to2$ differential cross section to calculate, and its calculation makes clear why it enjoys good high energy behavior.

\section{Optical Theorem}\label{s3}
Results for the high energy limit can be obtained by calculating for finite energy and mass, and then taking the $m\to0$ limit at the end of the calculation. Here we show that results in this limit can be obtained more directly. First we note the relation between the 4-derivative propagator,
\begin{align}
    G^{(4)}(p^2,m^2)=-\frac{1}{(p^2+i\varepsilon)(p^2-m^2+i\varepsilon)}
\end{align}
and the Feynman propagator,
\begin{align}
    G^{(2)}(p^2,m^2)=\frac{1}{p^2-m^2+i\varepsilon}
\end{align}
that occurs in the $m\to0$ limit. Given that 
\begin{align}
    G^{(4)}(p^2,m^2)&=-\frac{G^{(2)}(p^2,m^2)-G^{(2)}(p^2,0)}{m^2}
,\label{e5}\end{align}
we have
\begin{align}
    \lim_{m\to0}G^{(4)}(p^2,m^2)&=\lim_{m\to0}(-\frac{d}{dm^2})G^{(2)}(p^2,m^2).
\label{e3}\end{align}
In the Appendix we use the corresponding relation in position space to derive a short distance constraint on $G^{(4)}$.

In the derivation of the perturbative optical theorem, the imaginary part of a forward scattering amplitude ${\cal A}_{i\to i}$ is extracted by cutting propagators and using the relation
\begin{align}
    {\rm Im}(G^{(2)}(p^2,m^2))=-i\pi\delta(p^2-m^2)
.\end{align}
The analog of this for $G^{(4)}$ in the $m\to0$ limit is
\begin{align}
    \lim_{m\to0}{\rm Im}(G^{(4)}(p^2,m^2))&=-i\pi\lim_{m\to0}(-\frac{d}{dm^2})\delta(p^2-m^2)\label{e2}\\&=-i\pi\frac{d}{dp^2}\delta(p^2)
.\end{align}
The last equality shows the enhanced singularity for the imaginary part of the massless $G^{(4)}$ propagator. The second to last equality shows that we can use the propagator $G^{(2)}$  for the calculation of the imaginary part of the amplitude, and then in addition take an $m^2$-derivative with $m\to0$, for each propagator in a diagram.

This procedure can be carried over to the RHS of the optical theorem, involving the squared amplitudes and a sum over on-shell final states. The calculation proceeds normally except that an $m^2$-derivative with $m\to0$ is taken with respect to each final state particle, which requires that each such particle be assigned its own dummy mass $m_j$. The squared amplitude will depend on the values of these $m_j^2$'s via the on-shell conditions. Thus the term on the RHS of the optical theorem corresponding to $n$ $\phi$-particles in a final state $f$ takes the form
\begin{align}
    \lim_{m_j\to0}\left[\prod_{j=1}^n(-\frac{d}{dm_j^2})\right]|{\cal A}_{i\to f}(m_1,...m_n)|^2\label{e16}
.\end{align}
The final-state phase-space integral is implicit and the remaining dependence on the kinematic variable(s) of interest, such as the CM energy-squared $s$, is also implicit. For comparison, in the standard sum over final states we would, for each $\phi$, sum over the two mass eigenstates, with a relative minus sign due to the negative norm state, and with a $1/m^2$ factor due to the way the field is normalized. Thus we see how this is reproduced in the $m\to0$ limit by instead taking  $m^2$-derivatives.
\begin{figure}
    \centering
    \includegraphics[width=1\linewidth]{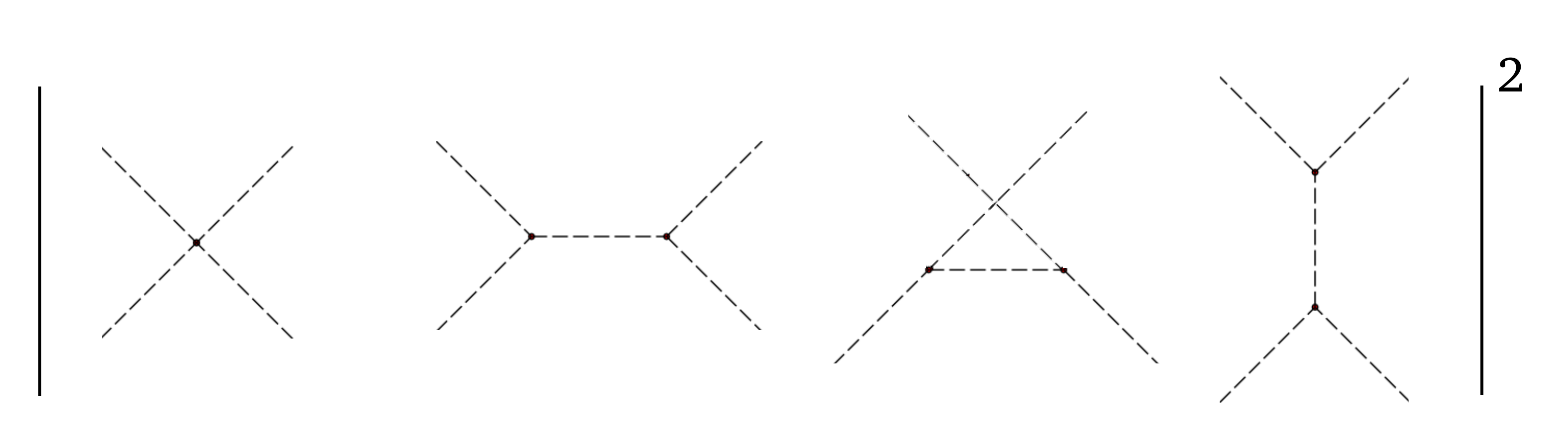}
    \caption{RHS of optical theorem involves this tree-level amplitude-squared with a sum over on-shell final states.}
    \label{f4}
\end{figure}
\begin{figure}
    \centering
    \includegraphics[width=1\linewidth]{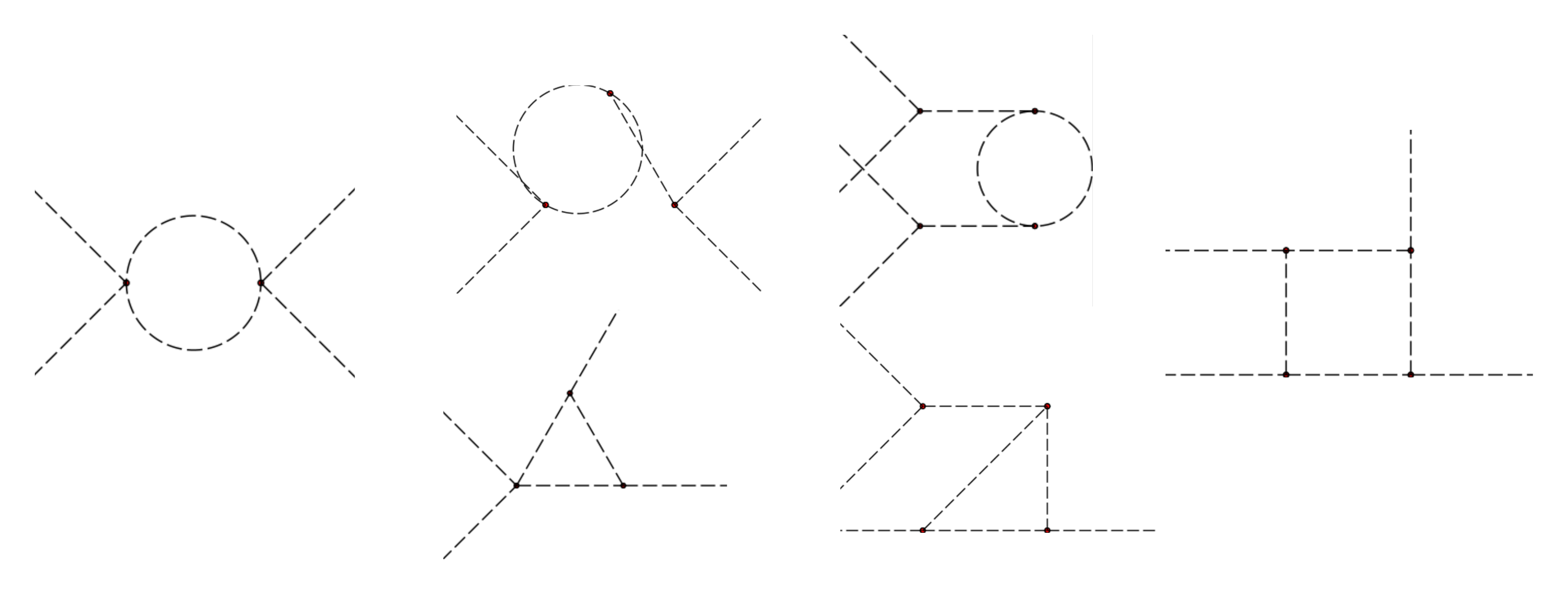}
    \caption{LHS of optical theorem involves the imaginary part of the forward scattering amplitude from these one-loop diagrams. Diagrams obtained by crossing are not shown.}
    \label{f3}
\end{figure}

Let us consider the optical theorem where the RHS describes the scattering of two of the massless particles into the possible final states via the amplitude-squared shown in Fig.~\ref{f4}. The LHS involves the set of diagrams as shown in Fig.~\ref{f3}, from which the imaginary part of the forward scattering amplitude is obtained. The various diagrams are of order $\lambda_4^2$, $\lambda_4\lambda_3^2$ or $\lambda_3^4$. The internal propagators can be represented as $m^2$-derivatives of Feynman propagators with $m\to0$. But infrared divergences are not present in the imaginary part, and so we can directly use the massless limit of the $G^{(4)}$ propagator,
\begin{align}
    G^{(4)}(p^2,0)=-\frac{1}{(p^2+i\varepsilon)^2}
.\end{align}
This square of a Feynman propagator can be in turn be handled by standard methods for calculating one-loop diagrams. Only the 4-derivative vertices of the theory remain a complication in the calculation.

The two high energy calculations are done independently and they give
\begin{align}
    \textrm{LHS}=\textrm{RHS}=\frac{s^2}{6\pi}(6\lambda_3^2+7\lambda_4)(\lambda_3^2+2\lambda_4)
\label{e10}.\end{align}
This confirms the S-matrix unitarity that we have argued is already expected. Note that the RHS naively has $s^4$ behavior, since the amplitudes go like $s^2$. This is reduced to $s^2$ behavior via the two $m^2$-derivatives.

\begin{figure}
    \centering
    \includegraphics[width=.85\linewidth]{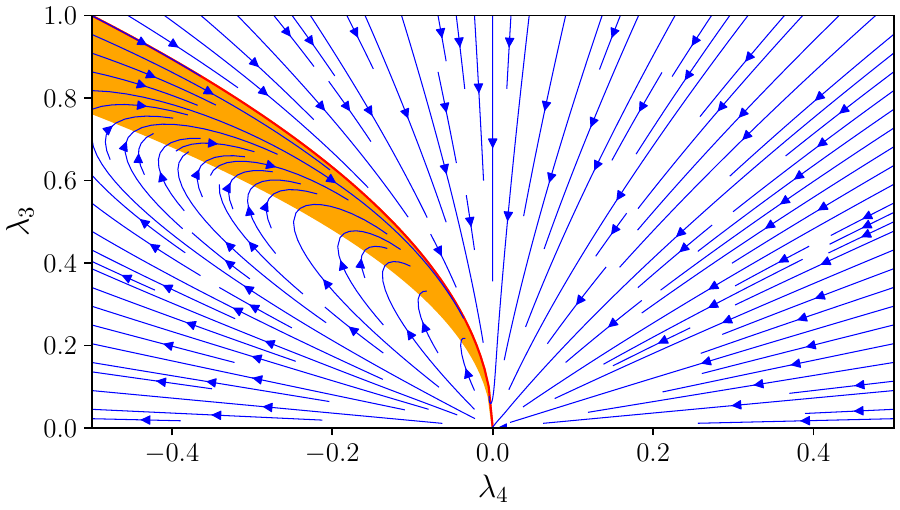}
    \caption{The LHS and RHS of the optical theorem becomes negative in the shaded region. Thus only the flows to the right of the red line are allowed.}
    \label{f1}
\end{figure}

The RHS of this optical theorem is, up to a positive normalization factor for the initial state, the total cross section for the scattering of the normal massless particles. This needs to be positive. While the squared amplitude is positive definite, this is no longer guaranteed after taking two $m^2$-derivatives. The result in (\ref{e10}) is indeed negative for $-\frac{6}{7}<\lambda_4/\lambda_3^2<-\frac{1}{2}$. We show this as the shaded region in Fig. \ref{f1}.
The upper boundary of this region is the same red line as in Fig.~\ref{f5}. The flows below this line will eventually enter the shaded region in the UV and thus all such flows are forbidden by the requirement of a positive cross section in the UV limit. The allowed flows correspond to the unshaded region in Fig.~\ref{f5}; these couplings are asymptotically free in the UV and become strong in the IR.

\section{A differential cross section}\label{s4}
Let us return to the question of the number of degrees of freedom in the high energy limit. Consider the two fields constructed from $\phi$,
\begin{align}
\psi_1&=\frac{1}{m^2}(\Box+m^2)\phi,\\
\psi_2&=\frac{1}{m^2}\Box\phi.
\end{align}
When expressed in terms of $\psi_1$ and $\psi_2$ the kinetic term of the Lagrangian becomes
\begin{equation}
    -\frac{m^2}{2}\psi_1\Box\psi_1+\frac{m^2}{2}\psi_2(\Box+m^2)\psi_2.
\label{e19}\end{equation}
Thus $\psi_1$ and $\psi_2$ are the two fields of definite mass ($0$ and $m$) and definite norm ($+$ and $-$). (The $m^2$ implies a $1/m^2$ normalization factor for external states that we have already noted.) But these fields are also such that $\phi=\psi_1-\psi_2$. All the interactions involving $\psi_1$ and/or $\psi_2$ are defined by interaction terms constructed from $\phi$, and thus the only combination that interacts is $\psi_1-\psi_2$. We are focused on the situation where the kinematic effect of the $\psi_2$ mass is infinitesimal in the high energy limit. The calculation of an amplitude can proceed by using the normal massive Feynman propagator (with a unique mass for each propagator) and then applying the multiple $-\lim_{m\to0}\frac{d}{dm^2}$ operations at the end. Similarly a differential cross section is calculated, after squaring the amplitude, by assuming each external line is a normal massive on-shell particle (with a unique mass for each external line) and then applying the multiple $-\lim_{m\to0}\frac{d}{dm^2}$ operations at the end. Thus these derivative operations are allowing the calculation to be performed as if there is just a single degree of freedom at high energies, in line with the observation that $\phi=\psi_1-\psi_2$.

We turn to calculating the differential cross section for $\phi\phi\to\phi\phi$, where this is now interpreted as the scattering of this single degree of freedom. This involves $m^2$-derivatives with respect to all four external lines, in contrast to the previous optical theorem calculation. We need the dependence on each mass $m_j$ associated with the $j$th external $\phi$, where these masses influence the amplitude through the on-shell constraints. The $m^2$-derivatives are to be taken with respect to the full $m_j$ dependence of the resulting differential cross section.

Before the $m^2$-derivatives are taken the result diverges as $\sim(s^2)^2/s$ for large $s$. This result is a complicated function of the $m_j$ and the scattering angle. A term like $\sim m_1^2m_2^2m_3^2m_4^2/s$ is needed to survive the four $m_j^2$-derivatives and the limit $m_j\to0$. We are then left with a differential cross section that behaves like $1/s$ at large $s$ times a function of the scattering angle. This provides a simple way to see why good high energy behavior is achieved. In \cite{Holdom:2023usn} we instead summed up exclusive cross sections, each behaving as $\sim(s^2)^2/s/m^8$ at high energies.

We find the following differential cross section for $\phi\phi\to\phi\phi$ scattering in the limit of high energies and as a function of both couplings,
\begin{align}
\frac{d\sigma}{d\Omega}=\frac{\left(\lambda_{3}^{4}-4 \lambda_{4}^{2}\right) \sin \! \left(\theta \right)^{6}+24 \lambda_{4}^{2} \sin \! \left(\theta \right)^{4}+\left(-48 \lambda_{3}^{4}-96 \lambda_{3}^{2} \lambda_{4} \right) \sin \! \left(\theta \right)^{2}+64 \lambda_{3}^{4}+128 \lambda_{3}^{2} \lambda_{4}}{16 \pi^{2} \sin \! \left(\theta \right)^{4} s}
.\label{e4}\end{align}
This result is positive definite for any value of $\theta$ as long as $\lambda_4\geq-\frac{1}{2}\lambda_3^2$. This is the same constraint that we found from the optical theorem that ruled out the shaded region in Fig.~\ref{f5}. We note that the $\sin(\theta)^{-4}$ pole is due to the $t$-channel exchange diagram. It is the coefficient of this $\sin(\theta)^{-4}$ pole that changes sign when crossing the red line in Fig.~\ref{f5}. Positivity is also achieved on the red line where $\lambda_4=-\frac{1}{2}\lambda_3^2$ and (\ref{e4}) reduces to $d\sigma/d\Omega=3\lambda_4^2/(2\pi^2s)$. Another special case has $\lambda_3=0$, in which case there is no exchange diagram and (\ref{e4}) again reduces to a positive result, $d\sigma/d\Omega=\lambda_4^2(5+\cos(\theta)^2)/(4\pi^2s)$.\footnote{The corresponding result eq.~(32) in \cite{Holdom:2023usn} has two typos, the 5 appeared in the wrong term and a factor of $1/64\pi^2$ is missing. The result also shows that the flow line with $\lambda_4<0$ and $\lambda_3=0$ is allowed, which becomes free in the IR instead of the UV.}

The constraint of positivity from two different calculations has picked out the running couplings that flow to strong coupling in the infrared. These calculations are probing the asymptotically free regime where the perturbative degrees of freedom are appropriate. This is similar to using perturbative QCD to study scattering of quarks and gluons at sufficiently high energies, even though quarks and gluons are not the asymptotic states. They are not asymptotic states because of strong interactions at intermediate energies. The 4-derivative theory also has strong interactions at intermediate energies, and so what are the asymptotic states in this theory?

We first note that the massless field $\psi_1=\frac{1}{m^2}(\Box+m^2)\phi$ transforms under the shift symmetry $\psi_1\to \psi_1+c$ in the same way as $\phi$ does, while $\psi_2$ does not transform. We can suppose that this shift symmetry is not broken by the strong interactions, and then $\psi_1$ can survive as a true asymptotic state with a normal massless kinetic term protected by symmetry. But the pole of the bare propagator for $\psi_2$ need not correspond to a true asymptotic state of the theory, due to strong interactions. The strong interactions might produce a spectrum of bound states, but even then any particular bound state need not correspond to an asymptotic state. An example of this from QCD is the sigma meson. Its width is of order its mass $\sim 0.5$ GeV, and so this is in no way an asymptotic state. The sigma meson is actually the lightest excitation in QCD other than the pseudo-Goldstone bosons, such as pions, that exist in the low energy theory. Its large width is due to its decay into pions.  Similarly, in the 4-derivative theory the massive ghost $\psi_2$ can decay into the $\psi_1$'s that exist in the low energy theory.

Whether it is through decay physics or something more like the confinement mechanism of QCD, we require that $\psi_2$ is not an asymptotic state. As long as all asymptotic states have positive norm, then all probabilities are positive. This does not mean that the ghost is invisible, since its virtual effects can be quite visible, for example by producing bumps in various energy distributions.

With this picture we end up with one propagating degree of freedom, at whatever energy scale is used to probe the theory. This is either $\psi_1$, the true asymptotic state, or $\psi_1-\psi_2$, the combination that appears in the perturbative high energy description. Our focus has been on high energies, where the ghost component plays an intrinsic role in achieving positivity and well-behaved cross sections. Positivity also happens to require that $\lambda_4\geq-\frac{1}{2}\lambda_3^2$, and this leads to the question of what is special about the boundary $\lambda_4=-\frac{1}{2}\lambda_3^2$.

When the couplings are related in this way and $m$ is identically zero, then the Lagrangian becomes a square ${\cal L}=-\frac{1}{2}(\Box\phi-\lambda_3\partial_\mu\phi\partial^\mu\phi)^2$.\footnote{I thank Neil Turok for pointing this out to me, as well as the related work in the next sentence.} This is then a theory of the form that has been studied in \cite{bogo,Rivelles:2003jd}, where it is argued that there is only a single physical state, the ground state. This relies on a symmetry that is present when $m=0$. In Section \ref{s3} we calculated the total cross section for the scattering of $\psi_1$ in the $m\to0$ limit. On the boundary $\lambda_4=-\frac{1}{2}\lambda_3^2$ our result in (\ref{e10}) vanishes. On the other hand in Section \ref{s4} we calculated the differential cross section for the $2\to2$ scattering of $\psi_1-\psi_2$, the state that is dictated by the perturbative theory. As we indicate below (\ref{e4}), on the boundary the result shows scattering that is isotropic but nonvanishing. We are finding that the $m\to0$ limit of the scalar theory on the boundary is particularly simple, but perhaps not quite as trivial as when working directly with the $m=0$ theory. Of course the scalar theory also has a $m\to0$ limit for the interacting theory away from the boundary.

In the gravity theory, QQG, the graviton and the spin-2 component of the massive ghost are analogous to $\psi_1$ and $\psi_2$. Only one combination of the two spin-2 fields should emerge in the perturbative description of ultra-Planckian scattering, and this should simplify the calculations of \cite{Holdom:2021hlo,Holdom:2021oii}. Any differential cross section involving the spin-2 fields (it can involve the lower spin components as well) turns out to be a function of a single coupling $Gm_G^2$ where $m_G$ is the ghost mass. This coupling is again asymptotically free, but since there is only one coupling involved, a positivity constraint on a 2d flow diagram does not arise here.

\section{Gauged extension}
In this section we consider another example of how 4-derivative kinetic terms can enable good high energy behavior. This is for an interacting $U(1)$ gauge field. The Stuckelberg representation of a massive $U(1)$ gauge field is
\begin{align}
    {\cal L}&=-\frac{1}{4g^2}F_{\mu\nu}F^{\mu\nu}+\frac{1}{2}m^2V_\mu V^\mu,\\&V_\mu=A_\mu+\partial_\mu\phi,\\
&F_{\mu\nu}=\partial_\mu V_\nu-\partial_\nu V_\mu=\partial_\mu A_\nu-\partial_\nu A_\mu.
\end{align}
If we try to add interaction terms, such as $\lambda_3V_\mu V^\mu\,\partial_\mu V^\mu+\lambda_4(V_\mu V^\mu)^2$, to this theory then we end up with amplitudes that grow like $s^2/m^4$ \cite{Kribs:2022gri}. So at most the interacting theory can only be a low energy effective theory. But our results for the 4-derivative scalar theory are suggesting that if we want to UV complete the Stuckelberg theory with interactions, then we need a 4-derivative kinetic term here as well. Thus we are led to the following gauged extension of our scalar theory,
\begin{align}
{\cal L}&=-\frac{1}{4g^2}F_{\mu\nu}F^{\mu\nu}+\frac{1}{2}V_\mu(\Box+m^2)V^\mu+\lambda_3V_\mu V^\mu\,\partial_\mu V^\mu+\lambda_4(V_\mu V^\mu)^2.\label{e23}
\end{align}
This is a starting point for a renormalizable theory of a $U(1)$ gauge-boson that is not only massive but is also interacting. We will again have to consider the issue of positivity.

Our previous shift symmetry has been promoted into a gauge symmetry, $\phi\to\phi-\alpha$ and $A_\mu\to A_\mu+\partial_\mu\alpha$, such that $V_\mu$ is invariant. We can add a gauge fixing term to the Lagrangian,
\begin{align}
    {\cal L}_\textrm{gf}=-\omega^*(\textbf{s}{\cal G})-\frac{1}{2\xi}{\cal G}^2.
\end{align}
$\textbf{s}$ is an operator on the space of fields that implements a BRST transformation and $\omega^*$ is the associated antighost. The four derivatives in the quadratic part of the action do not prevent us from defining a ${\cal G}$ that will realize a $R_\xi$-type gauge fixing. We generalize the discussion in \cite{Kribs:2022gri} to our case.

The choice
\begin{align}
    {\cal G}=\partial_\mu A^\mu-\xi(\Box+m^2)\phi
\end{align}
means that ${\cal L}_\textrm{gf}$ appears as the last two terms in the following quadratic part of the Lagrangian,
\begin{align}
     {\cal L}^{(2)}=&-\frac{1}{4g^2}F_{\mu\nu}F^{\mu\nu}+\frac{1}{2}(A_\mu+\partial_\mu\phi)(\Box+m^2)(A^\mu+\partial^\mu\phi)\nonumber\\&-\frac{1}{2\xi}\left(\partial_\mu A^\mu-\xi(\Box+m^2)\phi\right)^2-\omega^*\!\left(\partial^2-\xi(\Box-m^2)\right)\omega.
\end{align}
The terms mixing $A_\mu$ and $\partial_\mu\phi$ cancel to give 
\begin{align}
    {\cal L}^{(2)}=&-\frac{1}{4g^2}F_{\mu\nu}F^{\mu\nu}+\frac{1}{2}A_\mu(\Box+m^2) A^\mu-\frac{1}{2\xi}(\partial_\mu A^\mu)^2\nonumber\\&+\frac{1}{2}\partial_\mu\phi(\Box+m^2)\partial^\mu\phi-\frac{1}{2}\xi((\Box+m^2)\phi)^2-\omega^*\!\left(\partial^2-\xi(\Box-m^2)\right)\omega.\label{e8}
\end{align}
This leads to the photon propagator, in momentum space,
\begin{align}
    -i\langle0|T A_\mu A_\nu|0\rangle=-\frac{1}{p^2-\zeta \,m^{2}}\left(\zeta g_{\mu\nu}-\frac{\left( \zeta-\xi  \left(1-\zeta\right)\right) {p{_{\mu}}} {p{_{\nu}}}}{\left(p^2+\xi\left(p^2-m^{2} \right)  \right) }\right),\label{e11}
\end{align}
where $1/\zeta=1+1/g^2$, while the $\phi$ propagator is
\begin{align}
    -i\langle0|T\phi\phi|0\rangle=\frac{1}{{p^2} \left(m^{2}-{p^2} \right)-\xi  \left(m^{2}-{p^2} \right)^{2}}.\label{e12}
\end{align}

One can then obtain a combined propagator as
\begin{align}
    -i\langle0|T (A_\mu+\partial_\mu\phi) (A_\nu+\partial_\nu\phi)|0\rangle&=-i\langle0|T A_\mu A_\nu|0\rangle-i (ip_\mu)(-ip_\nu)\langle0|T\phi\phi|0\rangle\nonumber\\&=-\frac{1}{p^2-\zeta m^2}\left(\zeta g_{\mu\nu}+(1-\zeta)\frac{p_\mu p_\nu}{p^2-m^2}\right).
\label{e9}\end{align}
This combined propagator is independent of the gauge parameter, as expected due to the gauge invariance of $A_\mu+\partial_\mu\phi$. This indicates that there is actually no redundancy in the description, and the gauge symmetry could be considered ``fake'' \cite{Kribs:2022gri}. As for the physical parameter $\zeta$, it varies over $0\leq\zeta\leq1$ as $0\leq g^2\leq\infty$.

The combined propagator can also be written as 
\begin{align}
    -i\langle0|T (A_\mu+\partial_\mu\phi) (A_\nu+\partial_\nu\phi)|0\rangle=-\frac{\zeta}{p^2-\zeta m^2}\left(g_{\mu\nu}-\frac{p_\mu p_\nu}{p^2}\right)-\frac{p_\mu p_\nu}{p^2(p^2-m^2)}.
\label{e6}\end{align}
This can be compared to the usual decomposition of the standard massive gauge-boson propagator,
\begin{align}
    -\frac{1}{p^2-m^2}\left(g_{\mu\nu}-\frac{p_\mu p_\nu}{m^2}\right)=-\frac{1}{p^2- m^2}\left(g_{\mu\nu}-\frac{p_\mu p_\nu}{p^2}\right)+\frac{1}{m^2}\frac{p_\mu p_\nu}{p^2}.\label{e7}
\end{align}
The first term in each decomposition propagates the two transverse polarizations, and they only differ by appearance of $\zeta$ in (\ref{e6}). The second term in (\ref{e6}) represents the longitudinal polarization, but we see that it propagates two degrees of freedom rather than the one in (\ref{e7}). There are four instead of three degrees of freedom in total.

We have already learned how to deal with the two longitudinal degrees of freedom in the ungauged theory. This leads to an approach for calculation in the gauged theory. We can develop the Feynman rules for the theory in (\ref{e23}) by treating $A_\mu$ and $\phi$ as independent fields. There are internal and external lines for both fields, and vertices involving both fields as specified by (\ref{e23}). For the propagators, $-i\langle0|T A_\mu A_\nu|0\rangle$ can be set to the first term in (\ref{e6}), while $-i\langle0|T\phi\phi|0\rangle$ can be set to the second term, which is the propagator of the ungauged theory. These propagators also follow by setting the gauge parameter $\xi=0$ in (\ref{e11})  and (\ref{e12}). For each external $A_\mu$ line we sum the differential cross-section over the two transverse polarizations, with the square of each polarization 4-vector being $\zeta$. Each external $\phi$ line is treated as in the ungauged theory.

The actual physical external states in this theory are described by the combination of fields $A_\mu+\partial_\mu\phi$. Our calculational method means that various exclusive differential cross sections that involve various choices of $A_\mu$'s and $\phi$'s for the external lines are calculated separately and then added together to get the physical result.

\begin{figure}
    \centering
    \includegraphics[width=0.85\linewidth]{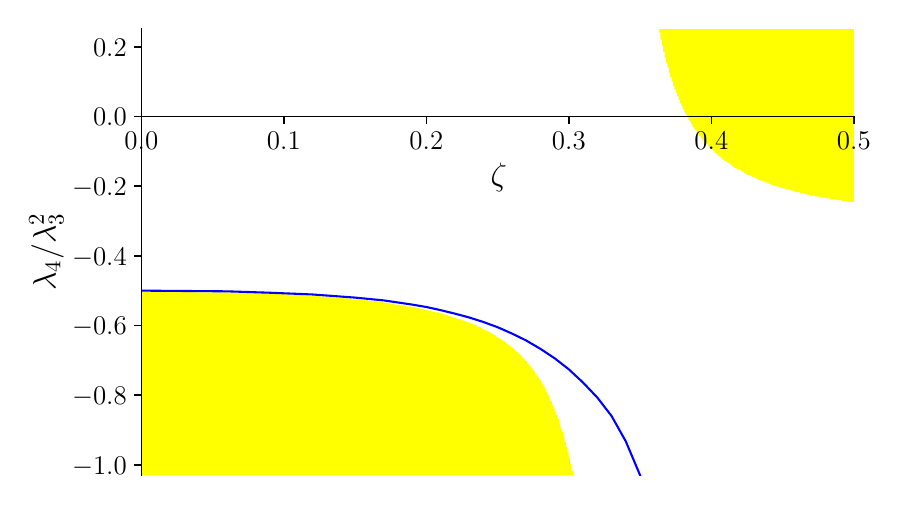}
    \caption{The unshaded region gives a $\phi\phi\to\phi\phi$ differential cross section with a positive $\sin(\theta)^{-4}$ pole. When positivity is required for the full physical cross section for any $\theta$, then $\lambda_4/\lambda_3^2$ must also be above the blue line. $\zeta$ grows with the gauge coupling as $1/\zeta=1+1/g^2$.}
    \label{f2}
\end{figure}

Let us consider tree-level $2\to2$ scattering in the high energy limit. We find that the various exclusive differential cross sections are separately well-behaved at high energies, all falling like $1/s$. The exclusive differential cross section for four $\phi$'s and no $A_\mu$'s in particular generalizes the one in (\ref{e4}), since it now picks up dependence on $\zeta$. This particular exclusive differential cross section is the only one that has the $\sin(\theta)^{-4}$ dependence, and near $\theta\approx0$ it is
\begin{align}
    \frac{d\sigma}{d\Omega}=\frac{4 {\lambda_3}^{2} \left(\left(\zeta^{2}-3 \zeta +1\right) {\lambda_3}^{2}-6 \zeta {\lambda_4} +2 {\lambda_4} \right)}{\pi^{2} \theta^{4}s}.\label{e14}
\end{align}
Thus the constraint that the $\theta^4$ pole in (\ref{e14}) be non-negative is a constraint that survives the sum over all the exclusive differential cross sections. With $f(\zeta)=(\zeta^2-3\zeta+1)/(6\zeta-2)$, the constraint is that
\begin{align}
    &\lambda_4/\lambda_3^2\geq f(\zeta) \quad\textrm{ for }\quad \zeta<1/3,\\
    &\lambda_4/\lambda_3^2\leq f(\zeta) \quad\textrm{ for }\quad \zeta>1/3.
\end{align}
This constraint is a restriction to the unshaded region in Fig.~(\ref{f2}). When $\zeta=0$ (zero gauge coupling) we recover the $\lambda_4/\lambda_3^2\geq-\frac{1}{2}$ constraint of the ungauged theory. As $\zeta$ increases, the constraint on $\lambda_4/\lambda_3^2$ gradually relaxes, $f(\zeta)$ becomes more negative, and it continues to relax up to the value of $\zeta=\frac{1}{3}$. Above this value there is instead an upper bound on $\lambda_4/\lambda_3^2$.

The physical differential cross section, the sum of all the exclusive ones, gives a complicated function of $\zeta$ and $\theta$. Although not enlightening, we can require that it be positive for any $0\leq\theta\leq\pi$, for each value of $\zeta$. This gives a further constraint on $\lambda_4/\lambda_3^2$, as shown by the blue line in Fig.~\ref{f2}. The allowed region lies above the blue line. 

The high energy limit in the gauged theory does not have quite the same significance as it does in the ungauged theory. This is because the $U(1)$ gauge coupling is, as usual, not asymptotically free and so this theory as a whole is not UV complete. Some new physics is needed before the Landau pole of the gauge coupling is reached. But this could be at an exponentially high energy scale, and our results apply at energies high compared to the mass but low compared to the new physics. Note that $\zeta>1/3$ corresponds to a quite large gauge coupling, and it may be that the required new physics has already entered before these values are reached.

We can also consider the low energy limit. At energies far below the mass there is again only a single massless scalar mode, since the gauge field is massive. Compared to the ungauged theory, there appears to be new contributions to the low energy theory coming from the exchange of the massive gauge-boson. But these exchange diagrams happen to vanish. Thus the low energy theory for the single massless scalar field, of both the gauged and ungauged theories, is the same, and is due to the exchange of the ghost.

\appendix*
\section{Propagator constraint}
Even though we can say little about the full $\phi$ propagator when there is a scale of strong interactions, we can obtain another constraint on the propagator in the asymptotically free regime at short distances. We start with the position-space version of the Feynman propagator $G^{(2)}$, using conventions where $x^2=x^\mu x_\mu>0$ is timelike,
\begin{align}
    G^{(2)}(x^2,m^2)=\theta(x^2)\frac{-im}{4\pi^2\sqrt{x^2}}K_1(im\sqrt{x^2})+\theta(-x^2)\frac{m}{4\pi^2\sqrt{-x^2}}K_1(m\sqrt{-x^2})+\frac{i}{4\pi^2}\delta(x^2).
\label{e17}\end{align}
Then by taking the $m^2$-derivative, that is the position-space version of (\ref{e3}),  we obtain the following result for $G^{(4)}$,
\begin{align}
    \lim_{m\to0}G^{(4)}(x^2,m^2)&=\lim_{m\to0}\left[-\theta(x^2)\frac{1}{8\pi^2}K_0(im\sqrt{x^2})-\theta(-x^2)\frac{1}{8\pi^2}K_0(m\sqrt{-x^2})\right]\\&\approx \theta(x^2)\frac{1}{8\pi^2}\left(\log(m\sqrt{x^2})+\frac{i\pi}{2}\right)+\theta(-x^2)\frac{1}{8\pi^2}\log(m\sqrt{-x^2})
.\label{e18}\end{align}
The last line assumes that $m\sqrt{x^2}\to0$, which is equivalent to $m/E\to0$ in the momentum-space propagator. Thus the quartic propagator has only a mild logarithmic singularity as the light-cone is approached from either side. This result is in turn related to the imaginary part of the momentum-space propagator. Or more precisely it is related to the energy contour used to calculate the position-space propagator from the momentum space propagator. We see that the mild logarithmic singularity in position space is a small distance constraint on the full propagator that is in addition to its $1/p^4$ behavior at large $p$.

\end{document}